\documentclass[a4paper,11pt]{article}
\usepackage{pos}
\usepackage{color}
\usepackage{amsmath, amsfonts, hyperref, color, graphicx, mathtools, empheq, academicons}
\newcommand{\di}{\text{d}}
\newcommand{\ddVt}{\tilde V^{(2)}}

\title{Analytic false-vacuum decay rate in the thin-wall approximation}

\author*[a,b]{Marco Matteini}
\author[a,b]{Miha Nemev\v{s}ek}
\author[a,c]{Lorenzo Ubaldi}

\affiliation[a]{Jo\v{z}ef Stefan Institute,\\
  Jamova 39, 1000 Ljubljana, Slovenia}

\affiliation[b]{Faculty of Mathematics and Physics, University of Ljubljana,\\
Jadranska 19, 1000 Ljubljana, Slovenia}

\affiliation[c]{Institute for Fundamental Physics of the Universe,\\
Via Beirut 2, 34014 Trieste, Italy}

\emailAdd{marco.matteini@ijs.si}
\emailAdd{miha.nemevsek@ijs.si}
\emailAdd{lorenzo.ubaldi@ijs.si}

\abstract{We present a fully analytical calculation of the false vacuum decay rate for a self-interacting scalar field in the thin-wall approximation. We obtain the bounce solution, together with the Euclidean action, counter-terms and renormalization group running, and we extract the functional determinant via the Gel'fand-Yaglom theorem. Our procedure is valid for a generic spacetime dimension $D$, and we provide an explicit finite renormalized decay rate in $D=3,4$. }

\FullConference{%
  Corfu Summer Institute 2022 "School and Workshops on Elementary Particle Physics and Gravity"\\
  28 August - 1 October, 2022\\
  Corfu, Greece
}

\begin{document}
\maketitle

\section{Introduction}\label{Sec_Intro}
Local ground states in physical systems may not be stable. A deeper minimum may exist, or appear with varying temperature, causing the system to undergo a phase transition that brings it from the metastable state, or “false vacuum”, to the stable state, or “true vacuum”. The order of the transition depends on the features of the system: a first-order phase transition is the most drastic case and consists of spherical regions (denoted as “bubbles”) of the true vacuum state nucleating within the false vacuum. If it is energetically favourable for the bubbles to grow, they will expand and convert the whole system to the true vacuum configuration \cite{Langer:1967ax,Coleman:1977py}. The probability of nucleating such bubbles is described by the “false vacuum decay rate” or “bubble nucleation rate”~\cite{Coleman:1977py,Callan:1977pt}. 

The topic of vacuum decay in high-energy physics is interesting in different contexts: the stability of the electroweak vacuum can be affected by unknown physics at arbitrarily large energy scales, thus modifying in a dramatic way the fate of our Universe. On the other hand, in the Early Universe, a first-order phase transition could have happened and would have left imprints on the cosmological history that might be visible today, such as a stochastic background of gravitational waves (GW). There is hope to detect such signals at present and future observatories, but at the moment the theoretical predictions for the power spectrum are afflicted by rather large uncertainties. 

One of the ingredients to determine the GW power spectrum is the bubble nucleation rate: it depends on the classical solution to the equations of motion of the decaying field, i.e. the bounce, via the bounce action, and on the quantum corrections. It is common in the literature to work with the effective potential in the exponential part of the decay rate, and estimate the dimensionful prefactor by dimensional analysis, based on the typical energy scale at which the transition takes place. In principle, this approximation might not be appropriate, but it is widely used due to the difficulty in dealing with the explicit calculation of the prefactor. 

In this work, I summarize the results obtained in \cite{Ivanov:2022osf} for the explicit and fully analytical calculation of the decay rate, including the prefactor, for a theory of a single real self-interacting scalar field at zero temperature, with a potential featuring almost-degenerate minima. We use the thin-wall approximation introduced in \cite{Coleman:1977py} to obtain analytical results expanded in powers of the thin-wall parameter, which specifies the difference in potential energy between the two minima.

This paper is organized as follows. In Section \ref{Sec_Setup} I set up the problem and the notation used throughout the paper. In Section \ref{Sec_Bounce} the expansion in the thin-wall parameter is introduced and the calculation of the bounce solution and the associated action, which enters into the exponential part of the decay rate, is performed. I also reproduce the procedure of dimensional regularization and renormalization in the $\overline{\text{MS}}$ scheme to take care of 1-loop divergences and calculate the counterterm action, which also enters into the exponential. In Section \ref{Sec_Fluctuations} I summarize the analytical procedure used in \cite{Ivanov:2022osf} to obtain the contribution coming from the quantum fluctuations. This procedure exploits the spherical $O(D)$ symmetry of the problem to perform an angular separation of variables, followed by an expansion in multipoles. We treat separately the low-multipole region, where the negative-eigenvalue mode and the zero modes are present, and the high-multipole region, which produces a result valid in the UV. In Section \ref{Sec_RenSum} the explicit contribution to the prefactor is obtained by performing the renormalized sum over the multipoles, and it is shown that the decay rate is free of divergences and independent of the renormalization scale. In Section \ref{Sec_Conclusion} I summarize our results and conclusions and present an outlook for future work.

\section{Setup of the problem}\label{Sec_Setup}
We consider a single real self-interacting scalar field $\phi$, with the potential
\begin{equation}
  V  = \frac{\lambda}{8} \left( \phi^2 - v^2 \right)^2 + \lambda \, \Delta \, v^3 \left(\phi - v \right)  \, .
\end{equation}
The first term is a mexican hat potential with degenerate minima. The linear term in $\phi$ breaks the degeneracy of the minima, and is proportional to the dimensionless parameter $\Delta$, which is the thin-wall parameter. The thin-wall limit of \cite{Coleman:1977py} is recovered for $\Delta \to 0$. Our results are built as a power-series expansion in $\Delta$, i.e., the thin-wall expansion. In our work, we consider $0 < \Delta \ll 1$ for the thin-wall expansion to be applicable, together with $0 < \lambda \ll 1$ to ensure perturbativity. The $\Delta$ parameter is chosen to be positive to keep the false vacuum on the right, and an upper bound is given by the disappearance of the false vacuum at $\Delta=1/\sqrt{27}$. Here the false vacuum and the maximum coincide, giving rise to an inflection point.

In quantum field theory the explicit 1-loop formula for the false vacuum decay rate can be obtained via the path integral formulation, and yields \cite{Coleman:1977py,Callan:1977pt}
\begin{equation}
 \frac{\Gamma}{\mathcal V} = \left( \frac{S_R}{2\pi \hbar} \right)^\frac{D}{2} \left\vert  
 \frac{\det' \mathcal O}{\det \mathcal O_\text{FV} }
 \right\vert^{- \frac{1}{2}} e^{-\frac{S_R}{\hbar} - S_{\rm ct} } \left( 1 + {\cal O}(\hbar) \right) \, ,
\end{equation}
for generic Euclidean spacetime dimension $D$. In the exponent, two terms appear: $S_R$ is the bounce action evaluated on renormalized couplings and $S_{\rm ct}$ is the 1-loop counterterm action. The prefactor contains the determinant of the fluctuation operator, explicitly given by $\mathcal O = -\partial_\mu \partial^\mu + d^2V/d\phi^2$, evaluated on the bounce solution in the numerator and on the false vacuum configuration in the denominator. The zero modes associated to the translational symmetry of the bounce are removed, hence the $\det^\prime$. There are $D$ such modes, and each removal introduces a factor of $\sqrt{S_R/(2\pi \hbar)}$, which appears in the prefactor above. As I will explicitly show, the $\log \mu$ and $1/\epsilon$ terms coming from $S_R$ and $S_{\rm ct}$ in the exponent are exactly cancelled by analogous terms coming from the regularized functional determinant. The $\log \mu$ is only present in even dimensions. In dimensional regularization, in four dimensions, we use the convention $\epsilon = 4-D$. 
In the above expression I kept the factors of $\hbar$ explicit, but from here on I will set $\hbar$ equal to 1.

\section{Bounce solution and action}\label{Sec_Bounce}
In flat spacetime, the bounce field configuration enjoys a spherical $O(D)$ symmetry \cite{Coleman:1977py}. The Euclidean action is given by
\begin{equation}
  S =\Omega \,  \int_0^\infty \text{d} \rho \, \rho^{D-1} \, \left( \frac{1}{2} \dot \phi^2 + V - V_{\rm FV}  \right) \, ,
\end{equation}
where $\Omega=2 \pi^{D/2}/\Gamma(D/2)$ is the surface element in $D$ dimensions, $\rho^2 = t^2 + x_i^2$ is the Euclidean radius, and the dot denotes a derivative with respect to $\rho$. The false vacuum contribution is explicitly subtracted in order to have a finite action. The equations of motion can be obtained by extremizing the action:
\begin{equation}
      \ddot \phi + \frac{D-1}{\rho} \dot \phi = \frac{\text{d} V}{\text{d} \phi} \, .
\end{equation}
The boundary conditions 
\begin{align}
\begin{split}
  \dot \phi(0) = \dot \phi(\infty) = 0\, ,
  \qquad
  \phi(0) &= \phi_{in} \, , 
  \qquad
  \phi(\infty) = \phi_{\text{FV}} \, ,
\end{split}
\end{align}
ensure the finiteness of the solution. At this stage, the initial value for the field is arbitrary. We will see that the thin-wall expansion forces it to be the true vacuum configuration.

\subsection{Thin-wall expansion}
We find it useful to switch to dimensionless variables, so we define the dimensionless field $\varphi$ and Euclidean coordinate $z$
\begin{equation}
    \varphi=\frac{\phi}{v} \, , \qquad z=\sqrt{\lambda}v\rho-r \, ,
\end{equation}
where the radius $r$ sets the size of the bubble wall.
When switching to dimensionless variables, we can factor out the dimensionful quantities in the action and obtain dimensionless equations of motion
\begin{align}
&S=\frac{\Omega v^{4-D}}{\lambda^{D/2-1}}\int_{-r}^\infty \text{d}z \, (z+r)^{D-1} \, \left( \frac{1}{2} \varphi'^2 + \Tilde{V} - \Tilde{V}_{\rm FV}  \right) \, , \\
&\varphi''+\frac{D-1}{z+r}\varphi'=\dfrac{\di\Tilde{V}}{\di\varphi} \, ,
\end{align}
where the prime denotes a derivative with respect to $z$ and $\Tilde{V}=V/(\lambda v^4)$.
We now expand the field and radius in powers of $\Delta$
\begin{equation}
    \varphi=\sum \varphi_n\Delta^n \, , \qquad r=\frac{1}{\Delta} \sum r_n\Delta^n\, .
\end{equation}
The expansion of the radius has an overall $1/\Delta$ factor, because the radius becomes infinite in the limit of degenerate minima, i.e., when $\Delta\to 0$.

The true and false vacuum configurations can be obtained as an expansion in $\Delta$ by requiring the derivative of the potential to vanish
\begin{align}
      &\varphi_{\rm TV} = - 1 - \Delta + \frac{3}{2} \Delta^2 - 4 \Delta^3 + {\cal O}(\Delta^4) \, , \\
      &\varphi_{\rm FV} = 1 - \Delta - \frac{3}{2} \Delta^2 - 4 \Delta^3 + {\cal O}(\Delta^4)  \label{bcphi} \, .
\end{align}

The bounce equation is now expanded in powers of $\Delta$ and solved order by order. Here I provide only the results. More details about the calculation can be found in \cite{Ivanov:2022osf}.

The first orders in $\Delta$ for the bounce are
\begin{align}
  \varphi_0 &= \text{th} \frac{z}{2} \, ,  
  \\  
   \varphi_1 &= -1 \, , \label{bounce1} 
   \\
  \begin{split}
    \varphi_2 & =  \frac{3}{4 (D-1) \text{ch}^2(z/2)} \bigl( 
    \left( 2 - D - 2 \left(4 + \text{ch} z \right) \ln(1 + e^z) \right) \text{sh} z
    \\
    & \qquad \qquad \qquad \qquad \qquad  - z \left(D - e^z \left(4 + \text{sh} z \right) \right) + 3 (\text{Li}_2(-e^z) - \text{Li}_2(-e^{-z}))
    \bigr) \, .
  \end{split}
\end{align}
The asymptotic behaviour
\begin{align}
  \varphi_0(z \to \pm \infty) &= \pm 1 \, ,
  &
  \varphi_1(z \to \pm \infty) &= -1 \, ,  
  &
  \varphi_2(z \to \pm \infty) &= \mp \frac{3}{2} \, ,
\end{align}
is consistent with the bounce approaching the false vacuum configuration at $z\to\infty$ (i.e., $\rho\to\infty$), and also shows that at $z\to-\infty$ (i.e., $\rho\to 0$) the bounce reaches the true vacuum. It is built in our construction that the bounce always interpolates between the false and the true vacuum. Pushing the expansion to higher orders in $\Delta$ simply increases the accuracy of our approximation.

At this stage we have fixed the shape of the bounce, but not its position. We need to fix the radius $r$. This can be done order by order either by solving the bounce equation at order $n+1$ and imposing the boundary conditions, or by extremizing the action at order $2n$
\begin{equation}
    \dfrac{\di S_{2n}}{\di r_n}=0 \, .
\end{equation}
The first three radii are
\begin{align}
  r_0 & = \frac{D-1}{3} \, , & 
  r_1 & = 0 \, , & 
  r_2 & = \frac{6 \pi^2 - 40 + D (26 - 4 D - 3 \pi^2)}{3(D-1)} \, .
\end{align}

We are now able to calculate the action. The leading order action is given by 
\begin{equation}
  S_0 = \frac{\Omega \, v^{4-D}}{\lambda^{D/2-1} \Delta^{D-1}} \left(\frac{D-1}{3} \right)^{D-1} \frac{2}{3 D} \, .
\end{equation}
This coincides with the one obtained in \cite{Coleman:1977py}. As expected, the action diverges in the thin-wall limit, suppressing the decay rate when the two minima are almost degenerate. The linear contribution in $\Delta$ vanishes, so the next-to-leading order is 
\begin{equation} \label{eq_action}
 S = S_0 \left( 1 + \Delta^2 \left(\frac{1 + D \left(25 - 8 D - 3 \pi^2 \right)}{2 (D-1)} \right) \right) \, .
\end{equation}
This new result gives a better approximation of the bounce action and also imposes an upper bound on $\Delta$, as the $\Delta^2$ term must be smaller than the leading order for the thin-wall expansion to be sensible. The contributions to the full action can be split in kinetic $\mathcal T$ and potential $\mathcal V $ contributions. Once this is done, it can be explicitly verified that Derrick's theorem $ \left(D - 2 \right) \mathcal T = -D \mathcal V $ is satisfied order by order in $\Delta$. This is a further check for the validity of our calculation.

\begin{figure}
    \centering
    \includegraphics[width=\textwidth]{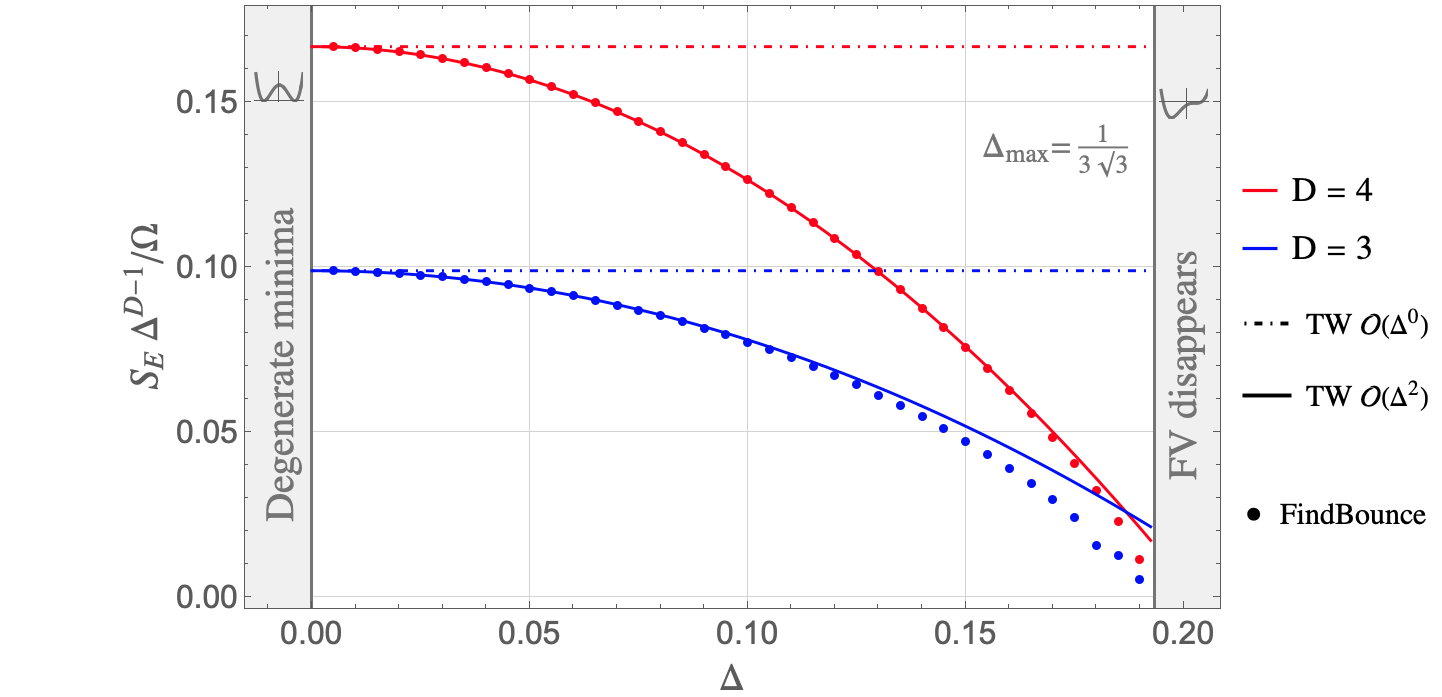}
    \caption{Comparison between the analytic result and the numerical result for the bounce action.}
    \label{fig:action}
\end{figure}

Explicitly, in $D=3$ the result is
\begin{equation}
   S=\frac{1}{\Delta^2} \frac{2^5  \pi v}{3^4 \sqrt{\lambda}} \left( 1 - \left( \frac{9 \pi^2}{4} - 1 \right) \Delta^2 \right) \, ,
\end{equation}
and in $D=4$
\begin{equation} \label{action4D}
S=\frac{1}{\Delta^3}\frac{\pi^2}{3 \lambda} \left( 1 - \left( 2 \pi^2 + \frac{9}{2} \right) \Delta^2 \right) \, .
\end{equation}
In figure \ref{fig:action} one can see the comparison between our analytic result and a numerical computation obtained via FindBounce \cite{Guada:2020xnz}. It is immediately clear that the inclusion of the $\Delta^2$ corrections is crucial in obtaining a decent approximate result away from the strict thin-wall limit, towards a situation where the false vacuum disappears and merges with the maximum of the barrier to become an inflection point. In an upcoming work \cite{Matteini:toappear} we investigate this further by including higher order corrections computed both analytically and semi-analytically.

\subsection{Renormalized action}
As we are calculating a 1-loop quantity, we need to take care of the renormalization of the action by introducing the relevant counterterms and performing the running of the couplings. We follow the usual dimensional regularization procedure and explicitly treat the $D=4$ case by introducing $\epsilon=4-D$.

We first consider the mexican hat potential with no linear term in $\Delta$ and then show that the inclusion of such term does not modify the 1-loop structure of the theory, i.e., the counterterm for $\Delta$ is zero and the parameter does not run. We define the counterterm potential, which will give the counterterm action, as
\begin{equation}
    V_\text{ct}=\frac{\delta_{m^2}}{2}\phi^2+\frac{\delta_\lambda}{4}\phi^4 \, .
\end{equation}
For simplicity of notation, we will write the $n$-th derivative of the potential as follows
\begin{equation}
      V^{(n)} \equiv \frac{\text{d}^n V}{\text{d} \phi^n} (\langle \phi \rangle) \, ,
\end{equation}
where $\langle \phi \rangle$ denotes the vacuum expectation value of $\phi$. We start by expanding the potential around $\langle \phi \rangle$. Removing the divergence of the four-point function fixes the quartic counterterm to
\begin{equation}
    \delta_\lambda=\frac{V^{(4)2}}{32\pi^2\epsilon} \, .
\end{equation}
The removal of the divergence in the three-point function is automatic if the three-vertex comes from the expansion of the quartic term in the potential around the vacuum expectation value. The removal of the divergence in the tadpole diagram fixes
\begin{equation}
    \delta_{m^2}=\frac{V^{(4)}}{16\pi^2\epsilon}\left( V^{(2)}-\frac{1}{2}V^{(4)}\langle\phi\rangle^2\right) ,
\end{equation}
and the cancellation of the divergence in the two-point function is automatic.

The above considerations are valid for a generic potential. When we turn on the linear term in $\Delta$, the third and fourth derivative do not change, therefore the quartic counterterm stays the same. The value of the vacuum expectation value $\langle \phi \rangle$ depends on $\Delta$, but the relations above hold for an arbitrary $\langle \phi \rangle$, therefore the introduction of the linear term does not modify our results, and the counterterm for $\Delta$ is zero.

The explicit counterterms for our potential are
\begin{equation}
    \delta_\lambda=\frac{9\lambda^2}{32\pi^2\epsilon} \, , \qquad \delta_{m^2}=-\frac{3\lambda^2v^2}{32\pi^2\epsilon} \, .
\end{equation}
The calculation of the 1-loop counterterm action can be explicitly performed in the thin-wall approximation, analogously to the calculation of the bounce action. The result is
\begin{equation} \label{CTaction}
    S_\text{ct}=\Omega \,  \int_0^\infty \text{d} \rho \, \rho^{D-1} \, \left(V_\text{ct}-V_\text{ct FV}\right)\simeq -\frac{3}{16\Delta^3\epsilon} \, .
\end{equation}

The running of the quartic coupling can be easily computed by requiring that the bare coupling 
does not depend on the renormalization scale $\mu$.
We obtain the following beta function
\begin{equation}
    \beta_\lambda=\frac{9\lambda^2}{16\pi^2}
\end{equation}
and
\begin{equation} \label{lambdarunning}
    \lambda(\mu)\simeq\lambda_0+\frac{9\lambda_0^2}{16\pi^2}\log\frac{\mu}{\mu_0} \, ,
\end{equation}
where $\lambda_0=\lambda(\mu_0)$ and the $\mu_0$ scale is arbitrarily chosen to be $\sqrt{V_\text{FV}^{(2)}}$.

We then plug in the running into the bounce action (\ref{eq_action}) and sum with the counterterm action to obtain the exponent of the decay rate
\begin{equation}
      S_R +  S_{\rm ct} = S \left( 1 -  \frac{9 \lambda_0}{16 \pi^2} 
  \left(\frac{1}{\epsilon} + \log \frac{\mu}{\mu_0} \right) \right) \, , 
\end{equation}
where $S$ is the tree-level bounce action from (\ref{eq_action}). We explicitly see the $1/\epsilon$ pole and the $\log\mu$ dependence that will be cancelled at 1-loop order by contributions coming from the prefactor. This is a non-trivial sanity check of the validity of our calculation.

\section{Fluctuations around the bounce}\label{Sec_Fluctuations}
Together with the analytical calculation of the higher order contributions of the bounce action, the most important result in our paper \cite{Ivanov:2022osf} is the analytical treatment of the ratio of determinants, which encodes the contribution from the fluctuations around the bounce. The two results combined give the fully analytical decay rate. In this section, I will briefly summarize the procedure which is thoroughly explained in \cite{Ivanov:2022osf}.

First of all, the spherical $O(D)$ symmetry of the problem is exploited to perform an angular separation of variables, followed by an expansion in multipoles $l$. The ratio of determinants then becomes 
\begin{equation}
 \left \vert \frac{\det' \mathcal O}{\det \mathcal O_\text{FV}}  \right\vert^{- \frac{1}{2}} = 
 \left\vert \prod_{l=0}^\infty \frac{\det'{\cal O}_l}{\det{\cal O}_{l{\rm FV}}}  \right\vert^{- \frac{1}{2}} \, ,
\end{equation}
with 
\begin{align}
  {\mathcal O}_l &= -\frac{\text{d}^2}{\text{d} \rho^2} - \frac{D-1}{\rho} \frac{\text{d}}{\text{d} \rho} + 
  \frac{l \left( l + D - 2 \right)}{\rho^2} + V^{(2)} \, , \label{eq:Ol}\\
  {\mathcal O}_{l{\rm FV}} &= -\frac{\text{d}^2}{\text{d} \rho^2} - \frac{D-1}{\rho} \frac{\text{d}}{\text{d} \rho} + 
  \frac{l \left( l + D - 2 \right)}{\rho^2} + V^{(2)}_{\rm FV} \, .\label{eq:OlFV}
\end{align}
The determinant of the operator is computed for fluctuations $\psi$ around the bounce $\overline \phi$. One way to compute the determinant would be to define a basis for $\psi$ and multiply the eigenvalues of the operator with fixed Dirichlet boundary conditions at $\rho = 0$ and $\infty$. Instead, we use the Gel'fand-Yaglom theorem~\cite{Gelfand:1959nq}, according to which
\begin{equation}
  \frac{\det {\cal O}_l}{\det {\cal O}_{l \rm{FV}}} = 
  \lim_{\rho \to \infty} R_l(\rho)^{d_l} \, ,
\end{equation}
where the degeneracy is given by~\cite{Kleinert:2004ev,Dunne:2006ct} 
\begin{align} 
  d_l &= \frac{(2l + D-2)(l+D-3)!}{l! (D-2)!} \, 
\end{align}
and
\begin{equation}
      R_l \equiv \frac{\psi_l}{\psi_{l{\rm FV}}} \, .
\end{equation}
The functions in the definition of $R_l$ are solutions of
\begin{align}
 \mathcal O_l \, \psi_l &= 0 \, , &
  \mathcal O_{l \text{FV}} \, \psi_{l\text{FV}} & = 0 \, ,
\end{align}
with fixed boundary conditions at the origin $\psi_{l,l\rm{FV}} (\rho \sim 0) \sim \rho^l$.

The ratio $R_l$ satisfies the differential equation 
\begin{equation}
    \ddot R_l + 2 \left( \frac{\dot \psi_{l\text{FV}}}{\psi_{l\text{FV}}} \right) \dot R_l =
  \left(V^{(2)} - V^{(2)}_\text{FV}   \right) R_l \, , 
\end{equation}
with boundary conditions $R_l (\rho = 0)= 1 \, ,  \dot R_l(\rho = 0) = 0$. We have reduced the problem of calculating a ratio of determinants to finding a solution to the differential equation for $\psi_{l\text{FV}}$, plugging it into the differential equation for $R_l$, and solving it. We chose to analyze $R_l$ directly~\cite{Baacke:2003uw}, because it is bounded on the entire $\rho$ interval thanks to the fact that $V^{(2)} - V^{(2)}_\text{FV} \xrightarrow{\rho \to \infty} 0$, which is not the case for $\psi_l$ and $\psi_{l\text{FV}}$, as they diverge exponentially when $\rho \to \infty$.

Already at this stage, before explicitly solving the differential equation for $R_l$, one can guess the following behaviour
\begin{align} \label{eq:Rlbehaviour}
  R_l(\infty) \, 
  \begin{cases}
    < 0, & l = 0 \, ,
    \\
   = 0, & l = 1 \, ,
    \\
    = 1, & l \gg 1 \, .
  \end{cases}
\end{align}
The ratio for the zero multipole should be negative to reflect the instability of the problem. For $l=1$, we expect the ratio to be zero due to the symmetry of the problem: the bubble can nucleate anywhere in the $D$-dimensional Euclidean spacetime, thus ensuring a translational symmetry. This is also reflected in the degeneracy factor: if we plug in $l=1$, we obtain exactly $d_l=D$. 

At very high multipoles, $R_l$ should go to $1$. The reason is that the only difference between $\psi_l$ and $\psi_{l\text{FV}}$ comes from the difference between $\mathcal O_l$ and $\mathcal O_{l \text{FV}}$ in \eqref{eq:Ol} and \eqref{eq:OlFV}, i.e., the second derivative of the potential evaluated on the bounce and on the false vacuum, respectively. In the high-multipole regime, this term is subdominant with respect to the $l$-dependent term.

\subsection{Low multipoles}
It is useful to introduce a new quantity 
\begin{equation}
    \nu=l-1+\frac{D}{2}
\end{equation}
that describes the multipoles. The reason for this shift is that, upon rescaling $\psi_{l{\rm FV}} \to \rho^{\frac{D-1}{2}} \psi_{l{\rm FV}}$ and switching to dimensionless variables, the friction term disappears from the differential equation of $\psi_{l{\rm FV}}$
\begin{equation}
    \frac{\text{d} \psi_{l\text{FV}}^2}{\text{d} z^2} = 
  \left(  \frac{\nu^2 - \frac{1}{4}}{(z+r)^2} + \ddVt_{\text{FV}}  \right) \psi_{l \rm{FV}}  \, , \quad  \ddVt_\text{FV} = \frac{1}{2} (3 \varphi_{\rm FV}^2 - 1) = 1 - 3 \Delta - 3\Delta^2 \, .
\end{equation}
We expand up to order $\Delta^2$ and drop higher order terms. The reason is apparent by looking at the differential equation
\begin{equation} \label{eq:psiFV}
     \frac{\text{d} \psi_{l\text{FV}}^2}{\text{d} z^2}
  = \left( 1 - 3 \Delta - 3 \Delta^2 + \Delta^2 \left( \frac{\nu^2 - \frac{1}{4}}{r_0^2} \right) \right) \, \psi_{l\text{FV}} \, .
\end{equation}
There is no multipole dependence at lower orders of $\Delta$. Conveniently, up to this order there is no $z$ dependence on the right hand side, so the solution can be easily found to be
\begin{equation}
      \psi_{l\text{FV}}(z)  \simeq c_{\rm FV} \exp \left[ \left( 1 - \frac{3}{2} \Delta + 
  \left(\frac{\nu^2 - \frac{1}{4}}{2 r_0^2} - \frac{21}{8} \right) \Delta^2 \right) z \right] \, .
\end{equation}
The false vacuum solution has the following form: $\psi_{l \text{FV}} = \psi_{l \text{FV}0} \psi_{l \text{FV}1}^\Delta \psi_{l \text{FV}2}^{\Delta^2} \ldots$. This suggests to perform an analogous multiplicative expansion for the ratio of determinants
\begin{align}
  R_l &= \prod_{n\geq 0} R_{l n}^{\Delta^n} \, ,
  &
  \log R_l = \sum_{n \geq 0} \log R_{l n} \Delta^n \, . 
\end{align}
This expansion is then inserted into the differential equation for $R_l$. I only show the final result here (more details can be found in \cite{Ivanov:2022osf})
\begin{equation}
  R_l(\infty) =  \Delta^2 e^{D-1} \frac{3}{4} 
  \frac{\left( l - 1 \right) \left( l + D - 1 \right)}{ \left( D - 1 \right)^2} \, .
\end{equation}
As expected, the dependence on the multipoles appears at order $\Delta^2$. Moreover, it can be immediately seen that this expression is negative for $l=0$ and vanishes for $l=1$, as anticipated in \eqref{eq:Rlbehaviour}. For high multipoles, the ratio does not go to $1$ at infinity, which is the expected behaviour in that regime. There is no need to worry: the above procedure is valid only for the low-multipole region.  In the differential equation for $\psi_{l\text{FV}}$, the multipole dependence enters via the $\Delta^2 \nu^2$ term, which was counted as $\mathcal O(\Delta^2)$. However, this power counting only makes sense as long as $\nu < 1 / \Delta$, which is not the case in the high-multipole regime.

\subsection{Zero removal}
The zero-eigenvalue modes ($l=1$) due to the translational invariance have to be removed from the product over the multipoles of the ratio of determinants. This can be done perturbatively~\cite{Jevicki:1976kd,Endo:2017gal,Endo:2017tsz,Andreassen:2017rzq} by off-setting the fluctuation operator by a small dimensionful parameter $\mu_\epsilon^2$ and solving
\begin{equation}
  \left( \mathcal O_1 + \mu^2_\epsilon \right) \psi_1^\epsilon = 0 \,     .
\end{equation}
The ratio of determinants then does not vanish and is given by 
\begin{equation}
     R^\epsilon_1(\infty) = \frac{\psi_1^\epsilon(\infty)}{\psi^\text{FV}_1(\infty)} 
   \simeq \frac{ \left(\mu^2_\epsilon + \gamma_1 \right) 
  \prod_{n=2}^\infty  \gamma_{\textbf n}}{\prod_{n=1}^\infty \gamma^{\text{FV}}_{\textbf n}}
  = \mu^2_\epsilon  R^\prime_1(\infty) \, ,
\end{equation}
where $\gamma$ are the eigenvalues and $\textbf n$ refers to the collective index over all the eigenvalues. The reduced determinant is thus explicitly given by
\begin{equation}
    R^\prime_1(\infty) = \lim_{\mu^2_\epsilon \to 0} \frac{1}{\mu^2_\epsilon} R^\epsilon_1(\infty) \, .
\end{equation}
Notice that the mass dimension of $R^\prime_1$ is reduced by 2 with respect to $R_l$. The details of the calculation can be found in \cite{Ivanov:2022osf}. Here I report the final result for the reduced determinant
\begin{equation}
    R^\prime_1(\infty) = \frac{e^{D-1}}{12} \frac{1}{\lambda v^2} \, 
\end{equation}
and the prefactor
\begin{equation}
     \left \vert \frac{\det' \mathcal O}{\det \mathcal O_\text{FV}}  \right\vert^{- \frac{1}{2}} = 
 \left( \left\vert R_0 \right \vert R_1^{\prime D} \prod_{l=2}^\infty \frac{\det{\cal O}_l}{\det{\cal O}_{l{\rm FV}}} \right)^{- \frac{1}{2}} \, .
\end{equation}
As expected, the removal of the zeroes gives the correct mass dimension to the decay rate.

\subsection{High multipoles}
The power counting to correctly treat the high-multipole region has to be $\Delta^2 \nu^2\sim\mathcal O(1)$. By looking at the right hand side of \eqref{eq:psiFV}, the new power counting causes the multipoles to appear already at leading order, instead of order $\Delta^2$. The leading order solution is thus given by 
\begin{align}
  \psi_{\nu \text{FV}} &\simeq e^{k_\nu z} \, ,
  &
  k_\nu^2 = 1 + \frac{\Delta^2 \nu^2}{r_0^2} \, .
  \label{knudefin}
\end{align}
We then move on to the leading order equation for the ratio $R$, whose solution at infinity can be shown to be 
\begin{align}
  R_{\nu 0}(\infty) &= \frac{\left( k_\nu - 1 \right)\left(2 k_\nu - 1 \right)}{\left( k_\nu + 1 \right)\left(2 k_\nu + 1 \right)} \, .
\end{align}
In \cite{Ivanov:2022osf} it was argued that this result alone is not complete: all orders in $\Delta^n$ give some contribution that feeds back into the leading order. Explicitly, this contribution amounts to
\begin{equation}
  U = 3 r_0 \left( k_\nu - \sqrt{k_\nu^2 -1} \right) \, .
\end{equation}
The full leading-order result is thus given by
\begin{equation} \label{eq:Rhighmultipole}
      \log R_\nu(\infty) = \log \frac{(k_\nu -1)(2k_\nu -1)}{(k_\nu+1)(2k_\nu+1)} + 3 r_0 \left( k_\nu - \sqrt{k_\nu^2 -1} \right) \, .
\end{equation}
In contrast to the low-multipole result, where the first non-vanishing term appeared at order $\Delta^2$, this is a leading order result in $\Delta$. As we will see, this difference will be crucial in the explicit computation of the sum over the multipoles:  the large multipoles dominate,  and at the leading order they are fully accounted for by~\eqref{eq:Rhighmultipole}. As expected from \eqref{eq:Rlbehaviour}, the determinant approaches 1 in the high-multipole regime, where $k_\nu \to \infty$.

\section{Computation of the renormalized sum}\label{Sec_RenSum}
Now the task is to perform the sum over the multipoles, which diverges for $\nu\to\infty$. The degeneracy factor is 
\begin{equation}
  d_\nu = \frac{2\nu \left( \nu + \frac{D}{2} -2 \right)!}{(D-2)! \left( \nu - \frac{D}{2} + 1 \right)!} \simeq 
  \frac{2\nu^{D-2}}{(D-2)!} \, .
\end{equation}
The approximation applies to the high multipole regime, but is an exact equality in $D=2, 3, 4$. Then we expand the high-multipole result for $R_\nu$, for example up to ${\cal O}(\nu^{-3})$, to obtain the explicit expression of the sum in this regime
\begin{equation}
  \sum_{\nu \gg 1} d_\nu  \log R_{\nu \gg 1} \sim -\frac{3 r_0 (2 - r_0)}{(D-2)! \Delta} \,
  \sum_{\nu \gg 1} \nu^{D-2} \left(\frac{1}{\nu} - \frac{1}{\nu^3} \left(\frac{r_0}{2 \Delta} \right)^2 \right) \, .
\end{equation}
The number of divergent terms depends on the spacetime dimension $D$: in $D = 2, 3$, only the first term diverges and gives a logarithmic divergence in $D = 2$ and a linear one in $D = 3$, while in $D = 4$ the first term gives a quadratic divergence and the second a logarithmic one. As we are dealing with 1-loop quantities, divergences are to be expected and must be regularized, and in fact they were already encountered in the renormalized bounce action. Of course the final result for the decay rate, being a physical quantity, must be finite, and therefore the divergences must cancel out. As anticipated, there should, and will, be an exact cancellation between the divergences coming from the prefactor and those present in the exponent. To ensure this, calculations must be carried on consistently, meaning that the renormalization scheme must be the same in the two calculations. In order to implement the $\overline{\text{MS}}$ scheme, we use the $\zeta$-function regularization scheme developed in \cite{Dunne:2006ct}, which 
was shown to be equivalent to $\overline{\rm MS}$.

The regularization of the UV-divergent sum is performed via asymptotic subtractions denoted by $R_\nu^a$. We introduce the following notation:
\begin{align}
\Sigma_D &= \sum_{\nu = \nu_0}^\infty\, \sigma_D = \sum_{\nu = \nu_0}^\infty d_\nu
  \left( \log R_\nu - \log R_\nu^a \right) \, ,
\end{align}  
where the lower boundary is given by $\nu_0 = D/2 -1$. To explicitly perform the sum, we use the Euler-Maclaurin (EuMac) approximation, which approximates the sum as an integral plus corrections from the boundaries:
\begin{align}
  &\Sigma_D \simeq \Sigma_D^{\int} + \Sigma_D^{\text{bnd}} + R_p \, ,
  \\
  &\Sigma_D^{\int} = \int_{\nu_0}^\infty \text{d} \nu \, \sigma_D \, ,\\
  &\Sigma_D^{\text{bnd}} = \frac{1}{2} \sigma_D(\nu_0)   - \sum_{j = 1}^{\left \lfloor {\frac{p}{2}}\right \rfloor} \frac{B_{2 j}}{(2 j)!}
  \sigma_D^{(2 j - 1)}(\nu_0) \, ,
\end{align}
where $\sigma_D^{(j)}$ stands for the $j$-th derivative of the summand. $B_{2 j}$ are the Bernoulli numbers and $R_p$ is the $p$-dependent remainder, which is the desired order of approximation. In principle there would be contributions from the upper boundary as well, but these vanish because the subtraction of the asymptotic
part guarantees that the sum is finite. In the following, I present the details of the explicit calculation for $D=3$ and $D=4$.

\subsection{Renormalized sum in $D=3$}
In $D=3$ only the first term in the sum diverges, therefore a single subtraction is sufficient, and the sum is given by~\cite{Dunne:2005rt, Dunne:2006ct} 
\begin{align}
  \log \left( \frac{\det \mathcal O}{\det \mathcal O_{\text{FV}}} \right) &= 
  \sum_\nu d_\nu \left(\log R_\nu - \frac{1}{2 \nu} I_1 \right) \, ,
\end{align}  
where $d_\nu = 2 \nu$ and 
\begin{align}
  I_1 &= \int_0^\infty \text{d} \rho \, \rho \left( V^{(2)} - V^{(2)}_\text{FV} \right) 
  \simeq - 3 \left( 2 - r_0 \right) \left(\frac{r_0}{\Delta}\right)=-\frac{8}{3\Delta} \, .
\end{align}
The last equality is valid only for $D=3$. To explicitly clarify the $\Delta$ dependence, it is useful to introduce $y = \Delta \nu/r_0$, which behaves as $\mathcal O(\Delta^0)$ in the high-multipole regime. In terms of this new variable, the asymptotic subtraction is given by $\log R_\nu^a=-2/y$ and the integral in the EuMac approximation is
\begin{align}
  \Sigma_3^{\int} &\simeq 2 \left(\frac{r_0}{\Delta} \right)^2 \int_{y_0}^\infty 
  \text{d} y \, y \left( \log R_{\nu} + \frac{2}{y} \right)
  = \frac{1}{\Delta^2} \frac{20 + 9 \ln 3}{27} \, .
\end{align}
This result does not depend on the precise value of the lower boundary of integration, which can then be extended to zero. In the integral we used the high-multipole result for $\log R_\nu$. The contribution from low-multipoles is subdominant, suppressed by $\Delta^2$, and we are only interested in the leading-order terms in this section.

The remaining terms depend on $\sigma_3(\nu_0)$, where we consider $\nu \sim \mathcal O(1)$ and expand in small $\Delta$. The leading term comes from the asymptotic part and goes as $\sigma_3 \simeq 8/(3 \Delta) + \mathcal O(\Delta^0)$, which is also suppressed. The $\nu$-derivatives are suppressed even stronger, therefore the Bernoulli terms are irrelevant.

The final result for the renormalized determinant in $D=3$ is thus
\begin{equation}
    \log \left( \frac{\det \mathcal O}{\det \mathcal O_{\text{FV}}} \right) =  \frac{1}{\Delta^2} \frac{20 + 9 \ln 3}{27} \, .
\end{equation}
As we are working in the $\overline{\rm MS}$ scheme, there are no counter-terms
and no running in $D=3$.

\subsection{Renormalized sum in $D=4$}
In $D=4$ we need two subtractions~\cite{Dunne:2005rt, Dunne:2006ct} to regularize the sum
\begin{align}
\begin{split}
  \log \left( \frac{\det \mathcal O}{\det \mathcal O_{\text{FV}}} \right) &= 
  \sum_\nu d_\nu \left(\log R_\nu - \frac{1}{2 \nu} I_1 + \frac{1}{8 \nu^3} I_2 \right) - \frac{1}{8} \tilde I_2 \, ,
\end{split}
\end{align}
where $d_\nu = \nu^2$. The $I_1$ integral has the same form as $D=3$ but evaluates to $-3/\Delta$ in $D=4$, while $I_2$ and $\tilde I_2$ are given by
\begin{align}
  I_2 &= \int_0^\infty \text{d} \rho \, \rho^3 \left( V^{(2) 2} - V^{(2) 2}_\text{FV} \right)
  \simeq - 3 \left( 2 - r_0 \right) \left(\frac{r_0}{\Delta}\right)^3 = -\frac{3}{\Delta^3} \, ,
  \\ 
  \begin{split}
  \tilde I_2 &= \int_0^\infty \text{d} \rho \, \rho^3 \left( V^{(2) 2} - V^{(2) 2}_\text{FV} \right)
  \left( \frac{1}{\epsilon} + \gamma_E + 1 + \log \left(\frac{\mu \rho}{2}\right) \right) 
  \\
  &\simeq I_2 \left( \frac{1}{\epsilon} + \gamma_E + \frac{5}{4} + 
  \log \left(\frac{\mu r_0}{2 \sqrt \lambda v \Delta}\right) \right) \, ,
  \end{split}
\end{align}
where the equalities are valid for $D=4$. The asymptotic subtraction from $I_1$ removes the quadratic divergence, while the one from $I_2$ removes the logarithmic divergence. The renormalized contribution $\tilde I_2$ is outside the sum and explicitly contains the $1/\epsilon$ pole and depends on the renormalization scale $\mu$. As I will explicitly show, these are exactly cancelled in the final result for the decay rate.

We again introduce the $y$ variable and start by calculating the integral contribution in the EuMac approximation
\begin{align}
  \Sigma_4^{\int} &\simeq \frac{1}{\Delta^3} \int_{y_0}^\infty \text{d} y \, y^2 \left( 
  \log R_\nu + \frac{3}{2 y} - \frac{3}{8 y^3} \right)
  = \frac{3}{8 \Delta^3} \left( \frac{9 - 4 \sqrt 3 \pi}{36} + \log 2 y_0 \right)\, .
\end{align}
As in $D=3$, the main contribution comes from high multipoles where $\Delta\nu\sim 1$, with the low multipole contribution being $\Delta$-suppressed. However, there is an explicit dependence on the lower boundary of integration, $y_0=\Delta\nu_0/r_0$, which comes from the asymptotic subtraction that behaves as $1/y^3$ and integrates to $\log y_0$, and is in the low-multipoles region. I will show how this contribution disappears in the final result.

In the corrections coming from the boundary terms, the summands $\sigma_D$ are expanded in small $\Delta$, with $\nu_0\sim 1$. The leading order part comes from the asymptotic subtraction that behaves as $1/y^3$, for which
\begin{equation}
    \sigma_4^{(j)}(\nu_0)=\frac{3(-)^{j+1}j!}{8\Delta^3\nu_0^{j+1}} \, .
\end{equation}
Since $\nu_0\sim 1$, when this is plugged into the sum of the boundary contribution, this diverges and the Bernoulli terms grow large.

To resolve this issue, we can introduce a splitting point $\nu_1$, such that ${\cal O}(1) = \nu_0 \ll \nu_1 < 1/\Delta$ and the sum is split
\begin{equation} \label{SumSplit}
\Sigma_D = \Sigma_D^{\rm low} + \Sigma_D^{\rm high} = \sum_{\nu = \nu_0}^{\nu_1} \sigma_D + \sum_{\nu = \nu_1+1}^\infty \sigma_D \, .
\end{equation}
The evaluation of the high part sum can be done as above for the integral part, while the boundary contribution is now suppressed thanks to $\nu_1\gg 1$, and the result is 
\begin{align}
  \Sigma_4^{\rm high} &
  = \frac{3}{8 \Delta^3} \left( \frac{9 - 4 \sqrt 3 \pi}{36} + \log 2 y_1 \right)\, .
\end{align}
The low part of the sum can be explicitly evaluated: up to $\nu_1 < 1/\Delta$ the leading piece in the summand  comes from the $1/y^3$ term in $\log R_\nu^a$, and $\Sigma_4^{\rm low}$ is given by the Harmonic number $H_{\nu_1}$, which can be expanded for $\nu_1 \gg 1$
\begin{align} \label{Sig4low}
  \Sigma_4^\text{low} = - \frac{3}{8 \Delta^3} \sum_{\nu = 1}^{\nu_1} \frac{1}{\nu} 
  = - \frac{3}{8 \Delta^3} H_{\nu_1} \simeq - \frac{3}{8 \Delta^3} \left( \log \nu_1 + \gamma_E \right) \, .
\end{align}
The low-multiple contribution to the sum is again $\Delta^2$-suppressed.
The final result for the regularized sum in $D=4$ is then 
\begin{align} \label{eqFuncDetD4}
  \Sigma_4 &  = \frac{3}{8 \Delta^3} \left( \frac{9 - 4 \sqrt 3 \pi}{36} - \gamma_E  + \log 2 \Delta \right) \, ,
\end{align}
which does not depend on the arbitrary splitting point $\nu_1$, and the renormalized determinant is
\begin{equation}
    \log \left( \frac{\det \mathcal O}{\det \mathcal O_{\text{FV}}} \right) =  \Sigma_4 - \frac{1}{8} \tilde I_2 = \frac{3}{8\Delta^3}\left(\frac{1}{\epsilon}+\log\left(\frac{\mu}{\sqrt{\lambda}v}\right) +\frac{3}{2}-\frac{\pi}{3\sqrt{3}}\right) \, .
\end{equation}

We can now explicitly check the cancellation of the $1/\epsilon$ pole and the $\log\mu$ dependence in the decay rate
\begin{equation}
  \log \frac{\Gamma}{\mathcal V} \ni - S_R - S_{\rm ct}
  - \frac{1}{2} \left( \Sigma_4 - \frac{\tilde I_2}{8} \right) \, .
\end{equation}

The renormalized determinant contribution is multiplied by $-1/2$ to account for the square root of the prefactor, so it amounts to 
\begin{equation}
 -\frac{3}{16\Delta^3\epsilon}  \, , \quad
  -\frac{3}{16\Delta^3}\log\mu  \, .
\end{equation}
The $1/\epsilon$ term in the action comes from the counterterm action (\ref{CTaction}), while the $\log\mu$ comes from the running of $\lambda$ (\ref{lambdarunning}) which enters into the renormalized action (\ref{action4D}), giving exactly
\begin{equation}
 +\frac{3}{16\Delta^3\epsilon}  \, , \quad
  +\frac{3}{16\Delta^3}\log\mu  \, .
\end{equation}

\section{Conclusion}\label{Sec_Conclusion}
I present here the final explicit results for the decay rate\footnote{Note that the result in $D=4$ is slightly different with respect to the one presented during the talk at the Workshop, i.e., the contribution from the functional determinant here contains a $54$ instead of a $45$. As highlighted in the acknowledgements of \cite{Ivanov:2022osf}, there was an error in our calculation, which was pointed out to us after the Workshop.}
\begin{align} \label{eqGamVSum}
  \frac{\Gamma}{\mathcal V} \simeq \left(  \left(\frac{S}{2 \pi}\right) \frac{12}{e^{D-1}}
  \lambda v^2 \right)^{D/2} \exp \left [ -S - \frac{1}{\Delta^{D-1}}
  \begin{cases}
    \frac{20 + 9 \ln 3}{54} \, , & D = 3 \, ,
    \\
    \frac{54 - 4 \pi \sqrt 3}{192}  \, , & D = 4 \, ,
  \end{cases} \right]
\end{align}
with the Euclidean bounce action given by
\begin{align} \label{eqSDelta2}
  S &= \frac{1}{\Delta^{D-1}}
  \begin{cases}
    \frac{2^5  \pi v}{3^4 \sqrt{\lambda}} \left( 1 - \left( \frac{9 \pi^2}{4} - 1 \right) \Delta^2 \right) \, , 	& D = 3 \, ,
    \\
    \frac{\pi^2}{3 \lambda} \left( 1 - \left( 2 \pi^2 + \frac{9}{2} \right) \Delta^2 \right) \, , & D = 4  \, .
  \end{cases}
\end{align}
As expected, the rate goes to zero both when $\lambda\to 0$ (the potential vanishes), and when $\Delta\to 0$ (the vacua are degenerate). Depending on the relative size of the couplings, the $\Delta^2$ correction of the bounce action and the 1-loop contribution from the functional determinant may be of comparable size and thus should both be included.

In this work I have summarized the results obtained in \cite{Ivanov:2022osf}. We studied the false vacuum decay process for a theory consisting of a self-interacting real scalar field with almost degenerate minima. We exploited the thin-wall expansion to obtain an analytical result for the bounce and for the bounce action up to next-to-leading order in the thin-wall parameter, and performed a 1-loop renormalization of the theory. The prefactor encoding the contribution from fluctuations around the bounce was treated using the Gel'fand-Yaglom theorem and exploiting the $O(D)$ symmetry of the problem to perform an expansion in multipoles. In the low-multipole region we performed the zero removal, which gives the correct physical dimension to the decay rate. At last, the sum over multipoles must be regularized in a consistent way, to ensure cancellation of divergences and unphysical dependencies with the corresponding terms in the renormalized action.

As mentioned, the immediate next step is pushing the thin-wall expansion to higher orders and check whether the series converges, and if so, if it approximates the numerical result better and better. This is studied in a work to appear \cite{Matteini:toappear}. Further generalizations of the methodology developed in \cite{Ivanov:2022osf} and summarized in the present work can explore different directions, such as a further analytical analysis and a numerical evaluation of the prefactor, and the inclusion of different fluctuations that couple to the decaying scalar field. These are examples of research directions towards a framework for evaluating the false vacuum decay rate at 1-loop for generic potentials, possibly with the help of a semi-analytical polygonal bounce setup \cite{Guada:2018jek}, which is already suited for multi-decaying-fields situations as well.

\acknowledgments
We gratefully acknowledge the contribution of Aleksandar Ivanov to the original work that this proceeding is based on. MM is supported by the Slovenian Research Agency’s young researcher program under grant No. PR-11241. MN is supported by the Slovenian Research Agency under the research core funding No. P1-0035 and in part by the research grants J1-3013, N1-0253 and J1-4389.


\begin{thebibliography}{99}

\bibitem{Langer:1967ax}
J.~S.~Langer,
\emph{Theory of the condensation point},
\href{https://doi.org/10.1016/0003-4916(67)90200-X}{Annals Phys. \textbf{41} (1967), 108-157}

\bibitem{Coleman:1977py}
S.~R.~Coleman,
\emph{The Fate of the False Vacuum. 1. Semiclassical Theory},
\href{https://doi.org/10.1103/PhysRevD.16.1248}{Phys. Rev. D \textbf{15} (1977), 2929-2936.}

\bibitem{Callan:1977pt}
C.~G.~Callan, Jr. and S.~R.~Coleman,
\emph{The Fate of the False Vacuum. 2. First Quantum Corrections},
\href{https://doi.org/10.1103/PhysRevD.16.1762}{Phys. Rev. D \textbf{16} (1977), 1762-1768.}

\bibitem{Ivanov:2022osf}
A.~Ivanov, M.~Matteini,  M.~Nemev\v{s}ek and L.~Ubaldi,
\emph{Analytic thin wall false vacuum decay rate},
\href{https://doi.org/10.1007/JHEP03(2022)209}{JHEP \textbf{03} (2022), 209.}
[arXiv:2202.04498 [hep-th]].

\bibitem{Guada:2020xnz}
V.~Guada, M.~Nemev\v{s}ek and M.~Pintar,
\emph{FindBounce: Package for multi-field bounce actions},
\href{https://doi.org/10.1016/j.cpc.2020.107480}{Comput. Phys. Commun. \textbf{256} (2020), 107480.}
 [arXiv:2002.00881 [hep-ph]].

\bibitem{Matteini:toappear}
M.~Matteini,  M.~Nemev\v{s}ek, Y.~Shoji and L.~Ubaldi, to appear.

\bibitem{Gelfand:1959nq}
I.~M.~Gelfand and A.~M.~Yaglom,
\emph{Integration in functional spaces and it applications in quantum physics},
\href{https://doi.org/10.1063/1.1703636}{J. Math. Phys. \textbf{1} (1960), 48.}

\bibitem{Kleinert:2004ev}
H.~Kleinert,
\emph{Path Integrals in Quantum Mechanics, Statistics, Polymer Physics, and Financial Markets}, WORLD SCIENTIFIC, 2009, \href{https://doi.org/10.1142/7305}{10.1142/7305.}

\bibitem{Dunne:2006ct}
G.~V.~Dunne and K.~Kirsten,
\emph{Functional determinants for radial operators},
\href{https://doi.org/10.1088/0305-4470/39/38/017}{J. Phys. A \textbf{39} (2006), 11915-11928.}
 [arXiv:hep-th/0607066 [hep-th]].

\bibitem{Baacke:2003uw}
J.~Baacke and G.~Lavrelashvili,
\emph{One loop corrections to the metastable vacuum decay},
\href{https://doi.org/10.1103/PhysRevD.69.025009}{Phys. Rev. D \textbf{69} (2004), 025009.}
[arXiv:hep-th/0307202 [hep-th]].

\bibitem{Jevicki:1976kd}
A.~Jevicki,
\emph{Treatment of Zero Frequency Modes in Perturbation Expansion About Classical Field Configurations},
\href{https://doi.org/10.1016/0550-3213(76)90403-X}{Nucl. Phys. B \textbf{117} (1976), 365-376.}

\bibitem{Endo:2017gal}
M.~Endo, T.~Moroi, M.~M.~Nojiri and Y.~Shoji,
\emph{On the Gauge Invariance of the Decay Rate of False Vacuum},
\href{https://doi.org/10.1016/j.physletb.2017.05.057}{Phys. Lett. B \textbf{771} (2017), 281-287.}
[arXiv:1703.09304 [hep-ph]].

\bibitem{Endo:2017tsz}
M.~Endo, T.~Moroi, M.~M.~Nojiri and Y.~Shoji,
\emph{False Vacuum Decay in Gauge Theory},
\href{https://doi.org/10.1007/JHEP11(2017)074}{JHEP \textbf{11} (2017), 074.}
[arXiv:1704.03492 [hep-ph]].

\bibitem{Andreassen:2017rzq}
A.~Andreassen, W.~Frost and M.~D.~Schwartz,
\emph{Scale Invariant Instantons and the Complete Lifetime of the Standard Model},
\href{https://doi.org/10.1103/PhysRevD.97.056006}{Phys. Rev. D \textbf{97} (2018) no.5, 056006.}
[arXiv:1707.08124 [hep-ph]].

\bibitem{Dunne:2005rt}
G.~V.~Dunne and H.~Min,
\emph{Beyond the thin-wall approximation: Precise numerical computation of prefactors in false vacuum decay},
\href{https://doi.org/10.1103/PhysRevD.72.125004}{Phys. Rev. D \textbf{72} (2005), 125004.}
[arXiv:hep-th/0511156 [hep-th]].

\bibitem{Guada:2018jek}
V.~Guada, A.~Maiezza and M.~Nemev\v{s}ek,
\emph{Multifield Polygonal Bounces},
\href{https://doi.org/10.1103/PhysRevD.99.056020}{Phys. Rev. D \textbf{99} (2019) no.5, 056020.}
[arXiv:1803.02227 [hep-th]].



\end{thebibliography}
\end{document}